\documentclass[manuscript]{acmart}
\AtBeginDocument{%
  \providecommand\BibTeX{{%
    \normalfont B\kern-0.5em{\scshape i\kern-0.25em b}\kern-0.8em\TeX}}}

\copyrightyear{2023}
\acmYear{2023}
\usepackage{amsfonts,amsmath,amsthm,bm}
\usepackage{booktabs}
\usepackage{graphicx,psfrag}
\usepackage{enumerate}
\usepackage{url}
\usepackage{algorithm}
\usepackage{algpseudocode}
\usepackage{lineno}
\usepackage{chngcntr}
\begin{document}
  \title{\bf Graph Encoder Ensemble for Simultaneous Vertex Embedding and Community Detection}
  \author{Cencheng Shen} 
  \email{shenc@udel.edu}
  \affiliation{%
  \institution{Department of Applied Economics and Statistics, University of Delaware}
  \city{Newark}
  \state{DE}
  \country{USA}
  \postcode{19716}
}

 \author{Youngser Park} 
  \email{youngser@jhu.edu}
  \affiliation{%
  \institution{Department of Applied Mathematics and Statistics, Johns Hopkins University}
  \city{Baltimore}
  \state{MD}
  \country{USA}
  \postcode{21218}
  }
  
   \author{Carey E. Priebe} 
  \email{cep@jhu.edu}
  \affiliation{%
  \institution{Department of Applied Mathematics and Statistics, Johns Hopkins University}
  \city{Baltimore}
  \state{MD}
  \country{USA}
  \postcode{21218}
}

\begin{abstract} %first sentence tweak
In this paper, we introduce a novel and computationally efficient method for vertex embedding, community detection, and community size determination. Our approach leverages a normalized one-hot graph encoder and a rank-based cluster size measure. Through extensive simulations, we demonstrate the excellent numerical performance of our proposed graph encoder ensemble algorithm.
\end{abstract}




\maketitle

\section{Introduction}
Graph data represents pairwise relationships between vertices through a collection of vertices and edges. Typically, a graph (or network) is represented by an adjacency matrix $\mathbf{A}$ of size $n \times n$, where $\mathbf{A}(i,j)$ denotes the edge weight between the $i$th and $j$th vertices. Alternatively, the graph can be stored in an edgelist $\mathbf{E}$ of size $s \times 3$, with the first two columns indicating the vertex indices of each edge and the last column representing the edge weight.

Community detection, also known as vertex clustering or graph partitioning, is a fundamental problem in graph analysis \cite{GirvanNewman2002, newman2004detecting, fortunato2010community, KarrerNewman2011}. The primary objective is to identify natural groups of vertices where intra-group connections are stronger than inter-group connections. Over the years, various approaches have been proposed, including modularity-based methods \cite{Louvain2008, Leiden2019}, spectral-based methods \cite{RoheEtAl2011, SussmanEtAl2012}, and likelihood-based techniques \cite{Gao2018, Abbe2018}, among others.

Spectral-based and likelihood-based methods are extensively studied in the statistics community, but they tend to be computationally slow for large graphs. On the other hand, modularity-based methods are faster and widely used in practice, but they lack theoretical investigations and only provide community labels without vertex embedding. Moreover, determining the appropriate community size poses a challenge for any method and is often addressed in an ad-hoc manner or assumed to be known. Therefore, a desirable approach is to develop a method that can achieve community detection, vertex representation, and community size determination under a unified framework.

In this paper, we propose a graph encoder ensemble algorithm that simultaneously fulfills all these objectives. Our algorithm leverages a normalized one-hot graph encoder \cite{GEE1}, ensemble learning \cite{opitz1999, Breiman2001}, k-means clustering \cite{Lloyd1982, Forgy1965}, and a novel rank-based cluster size measure called the minimal rank index. The proposed algorithm exhibits linear running time and demonstrates excellent numerical performance. The code for the algorithm is available on GitHub\footnote{\url{https://github.com/cshen6/GraphEmd}}.

\section{Methods}
We begin by introducing the one-hot graph encoder embedding from \cite{GEE1}, known for its computational efficiency and theoretical guarantees under random graph models. This embedding forms the foundation of our proposed ensemble method, outlined in Algorithm~\ref{alg3}. The ensemble algorithm incorporates crucial enhancements, including $L2$ normalization, the minimal rank index, and ensemble embedding, which are elaborated in the subsequent subsections.

\subsection{Prerequisite}
Given the graph adjacency matrix $\mathbf{A} \in \mathbb{R}^{n \times n}$ and a label vector $\mathbf{Y} \in {1,\ldots,K}^{n}$, we define $n_k$ as the number of observations per class, where
\begin{align*}
n_k = \sum_{i=1}^{n} 1(\mathbf{Y}_i=k)
\end{align*}
for $k=1,\ldots,K$. We construct the one-hot encoding matrix $\mathbf{W} \in \mathbb{R}^{n \times K}$ on $\mathbf{Y}$, then normalize it by the number of observations per-class. Specifically, for each vertex $i=1,\ldots,n$, we set
\begin{align*}
\mathbf{W}(i, k) = \frac{1}{n_k}
\end{align*}
if and only if $\mathbf{Y}_i=k$, and $0$ otherwise. The graph encoder embedding is then obtained by performing a simple matrix multiplication:
\begin{align*}
\mathbf{Z}=\mathbf{A} \mathbf{W} \in \mathbb{R}^{n \times K}.
\end{align*}
Each row $\mathbf{Z}(i, :)$ represents a $K$-dimensional Euclidean representation of vertex $i$. The computational advantage of the graph encoder embedding lies in the matrix multiplications, which can be efficiently implemented by iterating over the edge list $\mathbf{E}$ only once, without the need for the adjacency matrix \cite{GEE1}. In Algorithm~\ref{alg3}, we denote the above steps as
\begin{align*}
\mathbf{Z}=\operatorname{one-hot-emb}(\mathbf{E}, \mathbf{Y}).
\end{align*}

\subsection{Main Algorithm}
The proposed ensemble method is described in detail in Algorithm~\ref{alg3}. It can be applied to binary or weighted graphs, as well as directed or undirected graphs. Throughout this paper, we set the number of random replicates $r=10$, the maximum number of iterations $m=20$, and the clustering range is determined based on the specific experiment.

In the pseudo-code, the $L2$ normalization step is represented by $\mathbf{Z}=\operatorname{normalize}(\mathbf{Z})$, which normalizes each vertex representation to have unit norm (see Section~\ref{sec1} for more details). Additionally, given an embedding $\mathbf{Z}$ and a label vector $\mathbf{Y}$, the minimal rank index is denoted as $\operatorname{MRI}(\mathbf{Z}, \mathbf{Y}) \in [0,1]$, which measures the quality of clustering with a lower value indicating better clustering (details in Section~\ref{sec3}). The k-means clustering step is denoted as $\operatorname{k-means}(\mathbf{Z}, K)$, and the adjusted Rand index is denoted as $\operatorname{ARI}(\mathbf{Y}, \mathbf{Y}_2)$, which measures the similarity between two label vectors of the same size. The ARI is a popular matching metric that ranges from $-\infty$ to $1$, with a larger positive value indicating better match quality and a value of $1$ representing a perfect match \cite{Rand1971}.

\begin{algorithm}[H]
\caption{Graph Encoder Ensemble}
\label{alg3}
\begin{algorithmic}%[1]
\Require An edgelist $\mathbf{E}$, a range of potential cluster size $R$, number of random replicates $r$, and number of maximum iteration $m$.
\Ensure The graph embedding $\mathbf{Z} \in \mathbb{R}^{n \times \hat{K}}$ for all vertices, the estimated number of clusters $\hat{K}$, the cluster indices $\mathbf{Y} \in \mathbb{R}^{n}$, and the minimal rank index $ind \in [0,1]$.
\Function{Graph-Encoder-Ensemble}{$\mathbf{E}, R, r,m$}
\State $ind=1$; \Comment{initialize the index to pick best cluster size}
\For{$k \in R$}
\State $ind_2=1$; \Comment{initialize the index to pick best random replicate}
\For{$i = 1,\ldots,r$}
\State $\mathbf{\hat{Y}}_{k}=rand(k,n)$; \Comment{randomly initialize a label vector of length $n$ in $[k]$}
\For{$j = 1,\ldots,m$}
\State $\mathbf{\hat{Z}}_{k}=\operatorname{one-hot-emb}(\mathbf{E},\mathbf{\hat{Y}}_{k})$; 
\State $\mathbf{\hat{Z}}_{k}=\operatorname{normalize}(\mathbf{\hat{Z}}_{k})$; 
\State $\mathbf{\hat{Y}}_{k}^{'}=\operatorname{k-means}(\mathbf{\hat{Z}}_{k},k)$; 
\If{$\operatorname{ARI}(\mathbf{\hat{Y}}_{k}, \mathbf{\hat{Y}}_{k}^{'})$==1} break;
\Else $\mathbf{\hat{Y}}_{k}=\mathbf{\hat{Y}}_{k}^{'}$;
\EndIf
\EndFor
\State $\mathbf{\hat{Z}}_{k}=\operatorname{one-hot-emb}(\mathbf{E},\mathbf{\hat{Y}}_{k})$; 
\State $\mathbf{\hat{Z}}_{k}=\operatorname{normalize}(\mathbf{\hat{Z}}_{k})$; 
\State $ind_3 = \operatorname{MRI}([\mathbf{\hat{Z}}_{k},\mathbf{\hat{Y}}_{k}])$; 
\If{$ind_3 < ind_2$}
\State $\mathbf{\hat{Z}}=\mathbf{\hat{Z}}_{k}$; $\mathbf{\hat{Y}}=\mathbf{\hat{Y}}_{k}$; $ind_2=ind_3$;
\EndIf
\EndFor
\If{$ind_2 \leq ind$}
\State $\mathbf{Z}=\mathbf{\hat{Z}}$; $\mathbf{Y}=\mathbf{\hat{Y}}$; $\hat{K}=k$; $ind=ind_2$;
\EndIf
\EndFor
\EndFunction
\end{algorithmic}
\end{algorithm}

\subsection{Why Normalization}
\label{sec1}

\begin{figure*}[ht]%bhp]
\centering
\includegraphics[width=1.0\linewidth,trim={3cm 0cm 2cm 0cm},clip]{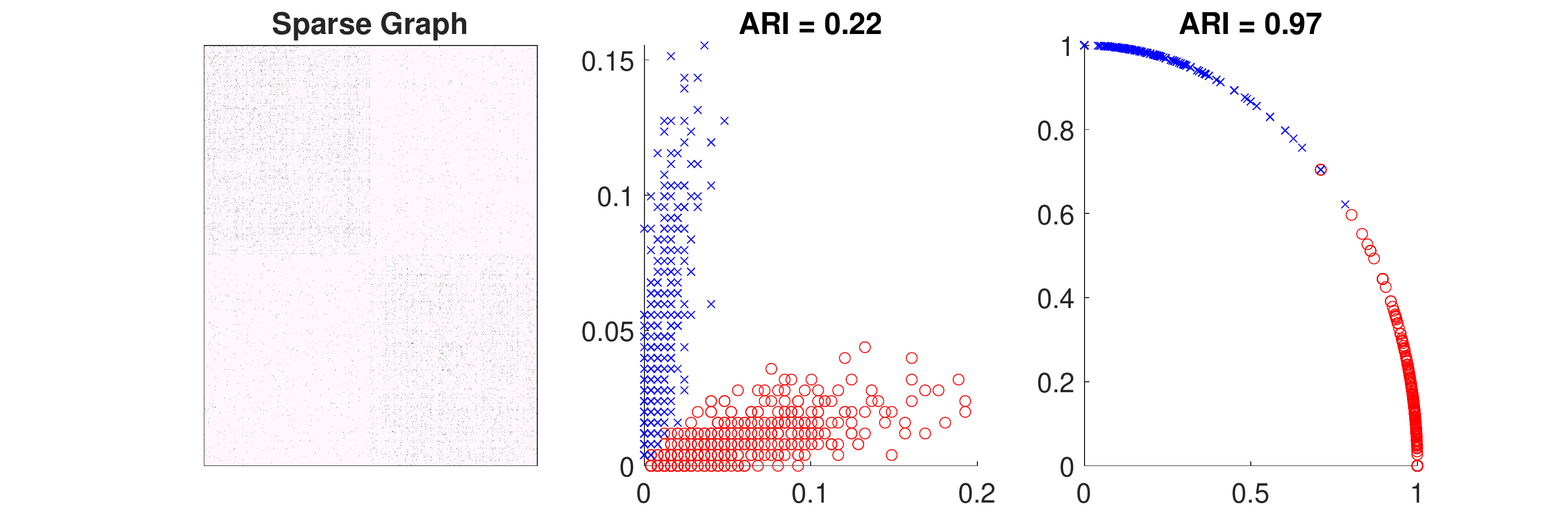}
\caption{This figure visually demonstrates the effect of normalization. The left panel displays the adjacency heatmap of a simulated sparse graph using simulation 1 in Section~\ref{sim1}. The center panel shows the resulting embedding without the normalization step, while the right panel displays the resulting embedding with normalization. The blue and red dots represent the true community labels of each vertex. }
\label{fig0}
\end{figure*}

The normalization step in Algorithm~\ref{alg3} scales each vertex embedding to unit norm. Specifically, for each vertex $i$, 
\begin{align*}
\mathbf{Z}(i,:)= \mathbf{Z}(i,:) / \|\mathbf{Z}(i,:)\|_{2}.
\end{align*}
if $\|\mathbf{Z}(i,:)\|{2} > 0$. The normalization step plays a crucial role in achieving improved clustering results, as demonstrated in Figure~\ref{fig0} using a sparse random graph model with two communities. The normalized embedding is represented on a unit sphere, effectively capturing the connectivity information while mitigating the influence of vertex degrees. In contrast, the un-normalized embedding is significantly affected by the original vertex degrees, resulting in vertices from the same community being widely dispersed. %Alternatively, using the normalized embedding with k-means clustering is equivalent to using the un-normalized embedding with angle, cosine, or spherical k-means clustering. 
This distinction bears resemblance to the two-truth phenomenon observed in graph adjacency and graph Laplacian, where the Laplacian spectral embedding (LSE) can be seen as a degree-normalized version of the adjacency spectral embedding (ASE). The LSE typically performs better on sparse graphs. Further numerical evaluations on the normalization effect can be found in Section~\ref{sim2} and Table~\ref{table1}.

\subsection{The Minimal Rank Index}
\label{sec3}

We introduce a new rank-based measure called the minimal rank index (MRI) to assess the quality of clustering. This measure plays a crucial role in Algorithm~\ref{alg3} as it enables the comparison of multiple embeddings generated from different initializations and community sizes.

Given the cluster index $\mathbf{Y}{i}$ of vertex $i$, the Euclidean distance function $d(\cdot,\cdot)$, and the mean of the $k$th cluster denoted as 
\begin{align*}
\mu_k = \sum_{i =1,\ldots,n}^{\mathbf{Y}_{i} = k} \mathbf{Z}_{i}, 
\end{align*}
the minimal rank index is computed as:
\begin{align}
\label{eq1}
MRI= \sum\limits_{i =1,\ldots,n} I\{ \arg\min\limits_{k=1,\ldots,K} d(\mathbf{Z}_{i}, \mu_k) \neq \mathbf{Y}_{i} \} / n \in [0,1].
\end{align}
The MRI measures how often the vertex embedding is not closest to its corresponding cluster mean. A smaller MRI value indicates better clustering quality, with MRI equal to $0$ indicating that every vertex is closest to its cluster mean. In the context of k-means clustering, MRI is non-zero when the k-means algorithm fails to converge.

In comparison to common cluster size measures such as Silhouette Score, Davies-Bouldin index, Variance Ratio Criterion, and Gap criterion \cite{Silhouettes, DBIndex}, MRI is rank-based rather than based on actual distances. These other measures compute ratios of within-cluster distances to between-cluster distances. If any of these measures were used in Algorithm~\ref{alg3} instead of MRI, the choice of cluster size would be biased towards the smallest possible value. This is due to the incremental nature of graph encoder embedding in Algorithm~\ref{alg3}, where the embedding dimension is equal to the community size $k$. Consequently, within-cluster distances become smaller for smaller values of $k$, resulting in a bias towards the smallest $k$ when using actual distance. 

\subsection{Ensemble Embedding and Cluster Size Determination}
\label{sec2}

Ensemble learning is utilized in Algorithm~\ref{alg3} to improve learning performance and reduce variance by employing multiple models. The approach can be summarized as follows: for each value of $k$ in the cluster range, we generate a set of vertex embeddings and community labels using random label initialization. The model with the smallest MRI is selected as the best model. In cases where multiple models have the same smallest MRI, the average embedding is used.

Additionally, among all possible choices of cluster size $k$, the best embedding with the smallest MRI is selected. If there are multiple embeddings with the same smallest MRI, the one with the largest $k$ is chosen. For instance, if the MRI values are $0,0,0,0.1,0.2$ for $K=2,3,4,5,6$, the graph encoder ensemble would select $\hat{K}=4$.

\subsection{Computational Complexity Analysis}
Algorithm~\ref{alg3} comprises several steps, including one-hot graph encoder embedding, k-means clustering, MRI computation, and ensembles. Let $n$ be the number of vertices and $s$ be the number of edges. At any fixed $k$, the one-hot graph encoder embedding takes $O(nk+s)$, k-means takes $O(nk)$, and the MRI computation takes $O(nk)$. Therefore, the overall time complexity of Algorithm~\ref{alg3} is $O(rm(n \max(R)+s)$, which is linear with respect to the number of vertices and edges. The storage requirement is also $O(n \max(R)+s)$. In practical terms, the graph encoder ensemble algorithm exhibits remarkable efficiency and scalability. Testing on simulated graphs with default parameters and $\max(R)=10$, it takes less than 3 minutes to process 1 million edges and less than 20 minutes for 10 million edges. 

\section{Results}

In this section, we conduct extensive numerical experiments to demonstrate the advantages of the graph encoder ensemble, as well as the individual benefits of normalization, ensemble, and MRI. We compare these approaches against benchmarks including the algorithm without normalization, without ensemble, with MRI replaced, and using adjacency/Laplacian spectral embedding. The performance is evaluated using the adjusted Rand index (ARI), which measures the degree of agreement between the estimated communities and the ground-truth labels.

\subsection{Simulation Set-up}
\label{sim1}

The stochastic block model (SBM) is a widely used random graph model for studying community structure \cite{HollandEtAl1983, SnijdersNowicki1997}. Each vertex $i$ is associated with a class label $Y_i \in \{1,\ldots, K\}$. The class label may be fixed a-priori, or generated by a categorical distribution with prior probability $\{\pi_k \in (0,1) \mbox{ with }  \sum_{k=1}^{K} \pi_k=1\}$. Then a block probability matrix $\mathbf{B}=[\mathbf{B}(k,l)] \in [0,1]^{K \times K}$ specifies the edge probability between a vertex from class $k$ and a vertex from class $l$. For any $i<j$,
\begin{align*}
\mathbf{A}(i,j) &\stackrel{i.i.d.}{\sim} \operatorname{Bernoulli}(\mathbf{B}(Y_i, Y_j)), \\
\mathbf{A}(i,i)&=0, \ \ \mathbf{A}(j,i) = \mathbf{A}(i,j).
\end{align*}

The degree-corrected stochastic block model (DC-SBM) \cite{ZhaoLevinaZhu2012} is a generalization of SBM to better model the sparsity of real graphs. Everything else being the same as SBM, each vertex $i$ has an additional degree parameter $\theta_i$, and the adjacency matrix is generated by
\begin{align*}
\mathbf{A}(i,j) \sim \operatorname{Bernoulli}(\theta_i \theta_j \mathbf{B}(Y_i, Y_j)).
\end{align*}
In our simulations, we consider three DC-SBM models with increasing community sizes. In all models, the degrees are generated randomly by $\theta_i \stackrel{i.i.d.}{\sim} Beta(1,4)$. 

Simulation 1: $n=3000$, $K=2$, $Y_i = \{1,2\}$ equally likely, and the block probability matrix is
\begin{align*}
\mathbf{B}=\begin{bmatrix}
0.5, 0.1\\
0.1, 0.5 
\end{bmatrix}.
\end{align*}

Simulation 2: $n=3000$, $K=4$, $Y_i = \{1,2, 3,4\}$ with prior probability $[0.2,0.2,0.3,0.3]$, and the block probability matrix is
\begin{align*}
\mathbf{B}=\begin{bmatrix}
0.9, 0.1, 0.1, 0.1\\
0.1, 0.7, 0.1, 0.1\\
0.1, 0.1, 0.5, 0.1\\
0.1, 0.1, 0.1, 0.3\\
\end{bmatrix}.
\end{align*}

Simulation 3: $n=3000$, $K=5$, $Y_i$ with equally likely prior probability, and the block probability matrix satisfies $\mathbf{B}(i,i)=0.2$ and $\mathbf{B}(i,j)=0.1$ for all $i=1,\ldots, 5$ and $j \neq i$.

\subsection{Normalization Comparison}
\label{sim2}
Table~\ref{table1} provides clear evidence of the superior clustering performance achieved by the normalized algorithm compared to the un-normalized algorithm. To isolate the impact of normalization, we set $r=1$ and assume the cluster size is known. The observed improvement aligns with the phenomenon observed between adjacency spectral embedding (ASE) and Laplacian spectral embedding (LSE), where LSE, being a normalized version of ASE, consistently outperforms ASE.

\begin{table}[H]
\renewcommand{\arraystretch}{1.3}
\centering
{\begin{tabular}{c||c|c|c|c|}
\hline
  \multicolumn{5}{c}{ARI} \\
 \hline
 & GEE & GEE no norm & ASE & LSE \\
\hline
Simulation 1 & $\mathbf{0.91}$ & $0.10$   &  $0.23$  & $\mathbf{0.91}$  \\
\hline
%Simulation 2 & $0.71$ & $0.17$     &  $0.27$  & $\mathbf{0.75}$  \\
%\hline
Simulation 2  & $\mathbf{0.73}$  & $0.08$   &  $0.12$  & $0.65$  \\
\hline
Simulation 3  & $\mathbf{0.78}$ & $0.06$    &  $0.17$  & $\mathbf{0.78}$ \\
\hline
\end{tabular}
\caption{This table demonstrates the advantage of normalization in the graph encoder ensemble. The "GEE" column refers to the graph encoder ensemble using Algorithm~\ref{alg3}, while "GEE no norm" indicates that normalization is not applied. The reported results are averages obtained from $100$ Monte Carlo replicates.}
\label{table1}
}
\end{table}

\subsection{Ensemble Comparison}
\label{sim3}

In this simulation, we assume a known cluster size and conduct $100$ Monte Carlo replicates to compare the performance of the ensemble algorithm ($r=10$) with the no-ensemble version ($r=1$). The results in Table~\ref{table2} clearly demonstrate the superiority of the ensemble algorithm: it achieves higher mean ARI and significantly reduces the variance compared to the no-ensemble version. Based on our empirical observations, the default choice of $r=10$ yields satisfactory results across our experiments. Additionally, if the graph size is sufficiently large and the community structure is well-separated, using a smaller value of $r$ or even $r=1$ is sufficient. This is evident in simulation 1 of Table~\ref{table2}.

\begin{table}[H]
\renewcommand{\arraystretch}{1.3}
\centering
{\begin{tabular}{c||c|c|}
\hline
  \multicolumn{3}{c}{Average ARI + std} \\
 \hline
 & GEE & GEE ($r=1$) \\%& GEE (using SS)  \\
\hline
Simulation 1 & $0.91 \pm 0.01$  & $0.91 \pm 0.01$    \\%& $0.91 \pm 0.01$ \\
\hline
Simulation 2 & $0.79 \pm 0.02$  & $0.72 \pm 0.09$    \\%& $0.79 \pm 0.01$ \\
\hline
Simulation 3 & $0.89 \pm 0.01$  & $0.79 \pm 0.12$    \\%& $0.89 \pm 0.01$ \\
\hline
\end{tabular}
\caption{This table assesses the advantage of the ensemble approach in the graph encoder ensemble. The reported results include the mean and standard deviation of the Adjusted Rand Index (ARI) obtained from $100$ Monte Carlo replicates.}
\label{table2}
}
\end{table}

\subsection{Cluster Size Estimation}
\label{sim4}

\begin{figure*}[ht]%[tbhp]
\centering
\includegraphics[width=0.9\linewidth,trim={2.5cm 0cm 2.5cm 0cm},clip]{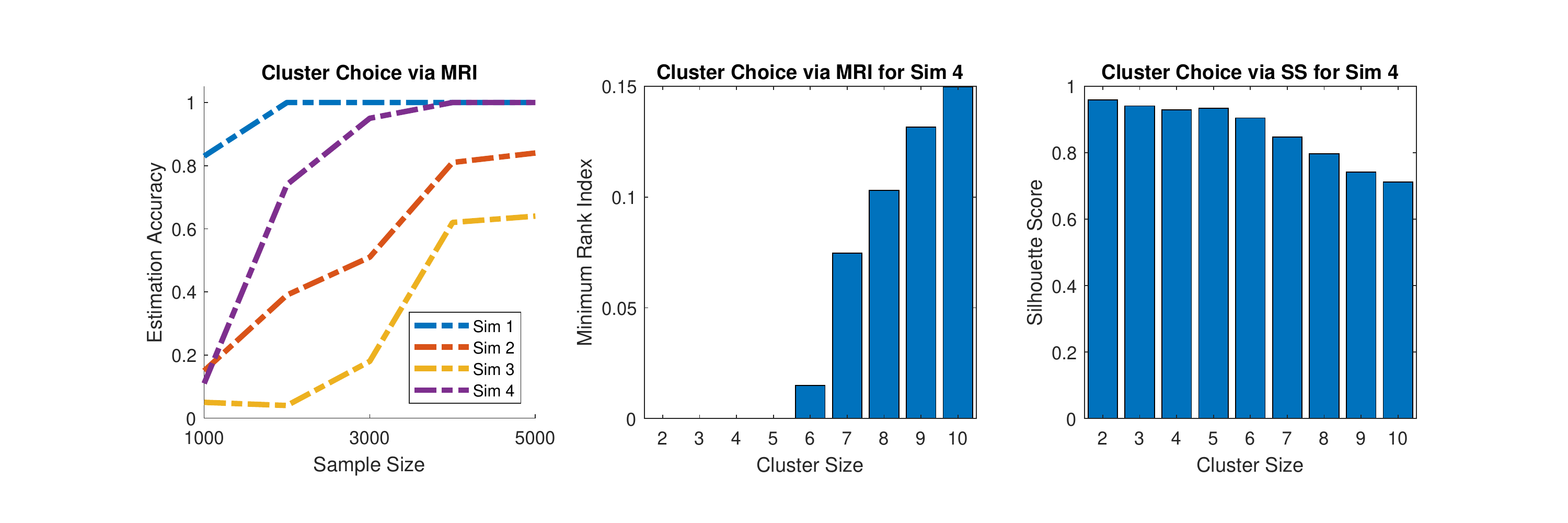}
\caption{This figure presents the results of cluster size estimation using the graph encoder ensemble. The estimation accuracy and the performance of different size measures are evaluated for various simulations and graph sizes. For each simulation and each graph size, we independently generate $100$ graphs, and run the ensemble algorithm to estimate the community size. The left panel of the figure illustrates the estimation accuracy as the graph size increases. The estimation accuracy represents the proportion of cases where the algorithm correctly chooses the community size. As the graph size increases, the estimation accuracy gradually improves, reaching a perfect estimation accuracy of $1$ for all simulations. The center panel focuses on simulation 3 at $n=5000$. The MRI calculates $\hat{K}=5$ as the estimated community size, which matches the ground-truth size. In the right panel, the average Silhouette Score is computed as an alternative size measure, which is biased towards smaller community sizes and chooses $\hat{K}_{SS}=2$, resulting in a different estimation compared to the ground-truth size.}
\label{fig1}
\end{figure*}

In this analysis, we explore the performance of the algorithm in estimating the community size. Instead of using the ground-truth size, we consider a range of potential sizes from $R=2$ to $R=10$, and the results are presented in Figure~\ref{fig1}.
These findings provide insights into the performance of the algorithm in accurately estimating the community size and highlight the importance of the MRI measure in achieving accurate size determination.

\section{Conclusion}
This paper introduces the graph encoder ensemble, which achieves graph embedding, community detection, and community size determination in a unified framework. Its main advantages include ease of implementation, computational efficiency, and excellent performance in community detection and community size selection. Several potential future directions include exploring mathematical proofs for asymptotic clustering optimality, investigating theoretical properties of MRI, and extending the method to dynamic and multi-modal graphs \cite{GEEFusion, GEEDynamic}.

\bibliographystyle{ACM-Reference-Format}
\bibliography{shen,general}

\begin{acks}
This work was supported in part by the National Science Foundation HDR TRIPODS 1934979, the National Science Foundation DMS-2113099, and by funding from Microsoft Research. 
\end{acks}

\end{document}